\documentstyle[11pt]{article}

\begin{document}

\begin{center}
  {\large\bf
Symmetry properties of the large-scale solar magnetic field
}
\end{center}

\begin{center}
  V. M. Galitski$^1$, K. M. Kuzanyan$^2$ and D. D. Sokoloff$^3$
\end{center}

\begin{center}
  {\it $^1$ Theoretical Physics Institute, University of Minnesota,
    55455 USA
\newline $^2$ IZMIRAN, Troitsk, Moscow Region,
      142190 Russia
\newline $^3$ Department of Physics, Moscow State
    University, 119899 Russia
}
\end{center}

{\footnotesize\noindent

We investigate symmetry properties of the solar magnetic field in the framework
of the linear Parker's migratory dynamo
  model. We show that the problem can be mapped onto
the well-known quantum-mechanical double-well problem. 
Using the WKB approximation, we construct analytical
  solutions for the dipole and quadrupole configurations of the
  generated magnetic field. 
Our asymptotic analysis within the equatorial region indicates
  the existence of an additional weak dynamo wave which violates the Hale's polarity law.
We estimate the spatial decay decrement of this wave.
We also calculate explicitly the splitting of the eigenvalues
  corresponding to the dipole and quadrupole modes. The real part of
  the dipole eigenvalue is shown to exceed the quadrupole one.
  A sufficiently long time after generation
the dipole mode should dominate the quadrupole one.
%
%
The relevant observational evidences of the
properties obtained are discussed.
}


\section{Introduction}

The problem of solar magnetic activity is an old and intriguing
subject.  A great amount of observational data have been accumulated
since the first half of 17th century.  A lot of efforts have been made to explain
theoretically these remarkable data.  However, the problem still bears
many unanswered questions. Among them, the problem of  symmetry of
the solar magnetic field exists.
%
%
This problem is visualized
as a symmetry violation in some tracers of the solar activity at low
latitudes in certain periods of times such as the Maunder Minumum
(Harvey, 1992; Ribes and Nesme-Ribes, 1993).

It is clear now that solar magnetic field generation is connected with
a turbulent motion of the differentially rotating electrically
conducting solar plasma.  
The full set of the corresponding
magnetohydrodynamic equations is too complicated to handle
analytically. However, a significant simplification arises if we limit
ourselves to considering the large-scale magnetic structure only. The
equations describing the large-scale magnetic field in a thin
convective shell were obtained by Parker on phenomenological grounds
back in 1955.  These equations have been intensively investigated for
almost fifty years.  A more solid basis of describing mean magnetic
field was provided by Steenbeck, Krause and R\"adler (1966) who
formally derived the equations of mean-field magnetohydrodynamics.
The Parker equations follow from the general equations of mean-field
magnetohydrodynamics being, however, only the leading approximation
with respect to the short length of a dynamo wave.  For a more
consistent treatment of the dynamo problem, the next-to-leading
approximation is required and the Parker equations must be slightly
modified (see {\em e.g.} the recent paper of Galitski and Sokoloff,
1999).

There is no doubt that the large-scale magnetic field
observed on the today's Sun corresponds to the steady regime.
Recent investigations of the Parker equations in the non-linear case have
revealed the following three features of the steadily oscillatory solutions
(see, {\em e.g.}, Meunier {\em et al.}, 1997; Bassom {\em et al.}, 1999;
Kuzanyan and Sokoloff, 1996). Firstly,
the structure of the steady state solution is not very sensitive to
the explicit form of the nonlinearity introduced into the mean-field
equations.  Secondly, the spatial profile of the dynamo wave in the
steady regime may retain some qualitative features of the one for the
kinematic problem. Thirdly, the frequency of the magnetic field
oscillations
coincides in the main approximation with the imaginary part of the
eigenvalue of the linear problem. Thus, it is quite reasonable to start
up with a relatively simple linear case in which the equations may
allow for an analytical solution.  Moreover, the kinematic problem is
a first step to approach the nonlinear case.

In the kinematic approximation, the mean-field equations reduce to an
eigenvalue problem for a linear differential operator. Formally, this
operator is quite similar to the standard Hamiltonian in
non-relativistic quantum mechanics. The diffusion terms play
the role of kinetic energy, while the
alpha-effect (helicity coefficient) and the
differential rotation play the role of a two-component external
potential.  It has been found fruitful to apply some
well-established quantum-mechanical approaches 
 to the problem at hand (e.g., Sokoloff {\em et al.},
1983, Ruzmaikin {\em et al.}, 1990).
Using the WKB approximation, Kuzanyan and Sokoloff (1995) and
later Galitski and Sokoloff (1998, 1999)
have obtained asymptotic analytical solutions of the
dynamo problem in the framework of the Parker model.  The solution
describing the wave of magnetic activity, the so-called dynamo wave,
was built up of the three
parts (dynamo waves): a strong dynamo wave propagating towards the
equator in the main spatial region, a relatively weak dynamo wave
propagating in the subpolar region polewards and an extremely weak
dynamo wave reflected from the pole and decaying exponentially
propagating equatorwards.  The former two waves are known from the
observations of the Sun (Makarov and Sivaraman, 1983),
while the latter wave predicted theoretically has not yet been
observed probably due to its weakness.

A major assumption adopted in the works cited above was that the
sources of magnetic field generation in the southern and northern
hemispheres were well separated and the generation took place in
different hemispheres absolutely independently.  Certainly, this
assumption is not appropriate when considering the symmetry problem in
which the interplay between the dynamo waves in different hemispheres
is the key factor. Moreover, the WKB approximation breaks down in the
very vicinity of the equator as shown below.

In the present paper, we investigate the symmetry problem by solving
the dynamo equations in the Parker's model. We investigate the
equatorial region and show that the wave of magnetic activity
generated in a given hemisphere does not vanish at the equator but
rolls over it propagating ahead in the other hemisphere and decaying
with propagation. Due to
this fact, there is a weak interaction between the dynamo waves in
different hemispheres which results in a small splitting of the
eigenvalues corresponding to different symmetry configurations of
the global magnetic field.
We estimate the spatial decay rate of this straying dynamo wave.
We also calculate the splitting explicitly and find
that the growth rate for the dipole configuration exceeds the one for
the quadrupole configuration in the framework of the Parker's model.
Notice, that the asymptotic method used enables us to estimate just
the asymptotic order of magnitude of this eigenvalue splitting and
we can hardly calculate the exact numerical coefficient by this method.

\section{Basic Equations}

In the present section we describe the model and establish the
necessary notations. Since, the current paper is based heavily on our
previous work, we highlight the main result only briefly omitting all
intermediate calculations. We refer an interested reader to the
original paper of Kuzanyan and Sokoloff (1995) [see also Galitski and
Sokoloff (1999)] for a more exhaustive presentation.

The starting point is the mean-field equations derived by Krause and
R\"adler (1980):
\begin{equation}
\label{MFE}
{\partial \bf B \over \partial t} =
{\bf \nabla} {\bf \times} \left( \alpha {\bf B} \right)+
{\bf \nabla \bf \times} \left( {\bf v \bf\times \bf B} \right) +
\beta \Delta {\bf B},
\end{equation}
where $\bf B$ and $\bf v$ are the large-scale (mean) magnetic and
velocity fields correspondingly, $\alpha$ is the helicity coefficient
and $\beta$ is the turbulent diffusivity.

In the linear approximation the time dependent part can be factored
out trivially:
\begin{equation}
\label{Bt}
{\bf B} ({\bf r},t) =  {\bf B} ({\bf r}) {\rm e}^{\gamma t},
\end{equation}
where $\gamma$ is an imaginary parameter to be found.

We consider the $\alpha\Omega$-dynamo model and suppose that magnetic
field generation occurs in a thin spherical shell corresponding to the
convective shell of the Sun. We also suppose that the differential
rotation is more intensive than the mean helicity and that the radial
gradient of the mean angular velocity $G = { \displaystyle 1 \over
  \displaystyle r} {\displaystyle \partial\Omega \over
  \displaystyle\partial r}$ weakly depends on latitude $\theta$.

To treat the problem, it is convenient to present the
axisymmetric magnetic field
as a superposition of the toroidal and poloidal components and express
the latter component as follows: ${{\bf B}_p} = {\rm \nabla \times} \,
\left( 0,\, 0,\, A \right)$ (we use the spherical system of coordinates
and measure off the latitude from the equator), where ${\bf A}$ a the
vector potential%
%
.

After averaging Eq.(\ref{MFE}) over the shell, one obtains
\begin{equation}
\gamma A = \alpha(\theta) B + {d \over {d
\theta}} \left[ {1 \over {\cos \theta}}
{d \over {d \theta}} (A
\cos \theta) \right],
\label{eqA}
\end{equation}
\begin{equation}
\gamma B =- D G(\theta) {d \over {d
\theta}} (A \cos \theta) + {d \over {d \theta}} \left[ {1 \over
{\cos \theta}} {d \over {d \theta}} (B \cos
\theta) \right].
\label{eqB}
\end{equation}
Note, that all quantities have been rescaled which leaded to the
simple dimensionless form of the equations. Parameter $D$ is the
dimensionless dynamo number which is supposed to be negative and
numerically large. Functions $\alpha(\theta)$ and $G(\theta)$ are
measured in units of their maximum values and they are certainly
unknown explicitly.
Observations of the solar magnetic activity and
helioseismological data (see {\em e.g.} Schou et al., 1998)
enable us to estimate them with a good accuracy ({\em e.g.},
Belvedere et al., 2000).
In what follows, we will suppose $\alpha(\theta)$ and $G(\theta)$
to be reasonable smooth functions subject to the following symmetry
constraints
$$
\alpha(\theta) = - \alpha(- \theta),\,\,\,\,\, \mbox{and}
\,\,\,\,\, G(\theta) = G(-\theta).
$$
One of the advantages of the asymptotic method we apply is that the
explicit knowledge of the functions is not required and the final
results can be expressed in a quite general form. To make explicit
estimates, we use the following simple form of the functions:
$\alpha(\theta) = \sin{\theta}$ and $G=1$.

It is convenient to rewrite Eqs.(\ref{eqA}, \ref{eqB}) in the following
symbolic form:
\begin{equation}
\label{spform}
\hat {\cal H} \vec h = \gamma \vec h,
\end{equation}
where we introduced a two-component complex function
\begin{equation}
\label{vech}
\vec h(\theta) = {A(\theta) \choose B(\theta)}
\end{equation}
and a matrix differential operator $\hat {\cal H}$ which is well-defined by
Eqs.(\ref{eqA}, \ref{eqB}):
\begin{equation}
\label{H}
\hat {\cal H} =
\left(
\begin{array}{cc}
\Delta&
\alpha(\theta) \\
\,\, & \,\,\\
-D G(\theta)
{\displaystyle d \over {\displaystyle d\theta}} \cos{\theta}&
\Delta
\end{array}
\right),
\end{equation}
where a notation for the azimuthal part of the Laplacian is
introduced for brevity $\Delta = {\displaystyle d \over {\displaystyle
    d\theta} } {\displaystyle 1 \over \displaystyle\cos{\theta} }
{\displaystyle d \over {\displaystyle d\theta} } \cos{\theta}$.  Note
that operator (\ref{H}) is nonhermitian and, thus, its
eigenvalues are complex in general.
The adjoint operator has the
following form
\begin{equation}
\label{Hd}
\hat {\cal H}^{\dagger} =
\left(
\begin{array}{cc}
\Delta&
D G(\theta)
{\displaystyle d \over {\displaystyle d\theta}} \cos{\theta}\\
\,\, & \,\,\\
\alpha(\theta) &
\Delta
\end{array}
\right).
\end{equation}
Note that operators $\hat{\cal H}$ and $\hat {\cal H}^{\dagger}$ are
defined on the segment $\theta \in \left[ -\pi/2, \pi/2\right]$.

The main idea of the WKB solution obtained in the above cited papers
was to expand both the eigenvectors and eigenvalues into an asymptotic
series in terms of parameter $\varepsilon = \left| D
\right|^{-1/3}$, which was supposed to be small.  Note, that magnetic
field generation occurs only due to the helicity and differential
rotation mechanisms [functions $\alpha(\theta)$ and
$G(\theta)\cos(\theta)$]. Remarkably, the two physically different
mechanisms collapse into one universal function $\tilde\alpha(\theta) =
\alpha(\theta) G(\theta) \cos(\theta)$ in the framework of the
asymptotic solution. This function is relatively small near the
equator and the poles. The magnetic field appears in these regions
mainly due to the diffusion from middle latitudes where the generation
is efficient.  The solution was obtained under the assumption that the
magnetic field generated in different hemispheres independently. This
is true if the interplay between the southern and northern dynamo
waves is negligible.

In the framework of the WKB approach, we present the solution $\vec n$
in the northern hemisphere as a product of a fast oscillating exponent
and a slowly varying amplitude $\vec n_0$:
\begin{equation}
\label{as}
\vec n(\theta) =
{\vec n_0(\theta)} {\rm exp} \left[ {i S(\theta) \over \varepsilon}\right].
\end{equation}

The eigenvalues are seeking in the form of the asymptotic series
\begin{equation}
\label{gamma}
\gamma = \varepsilon^{-2} \Gamma_0 + \varepsilon^{-1} \Gamma_{1,n},
\end{equation}
where integer index $n$ parameterizes the discrete eigenvalues.

Using the WKB approximation technique, one can obtain the following
explicit expression for the amplitude
\begin{equation}
\label{n0}
\vec n_0 (\theta) =
{1\over \cos{\theta}}
{\Gamma_0 + k(\theta)^2 \choose
i \varepsilon^{-2} k(\theta) \cos{\theta}} {\sigma(\theta)}
,
\end{equation}
where $k(\theta) = d S(\theta) / d \theta$ which satisfies the
following Hamilton-Jacobi equation:
\begin{equation}
\label{HJ}
\left[ \Gamma_0 + k^2(\theta) \right]^2 - i \tilde\alpha(\theta) k(\theta) =0,
\end{equation}
where $\tilde\alpha = \alpha(\theta) G(\theta) \cos{\theta}$.  This
algebraic equation possesses four branches of roots. It is impossible
to construct a smooth solution using any individual branch. Such a
solution appears only by matching two different branches.  A
continuous crossover from one branch to the other is possible only for
some unique values of the spectral parameter. This condition pins the
value of $\gamma$ and determines the desired spectrum:
\begin{equation}
\label{G0}
\Gamma_0= {3 \over 2^{8/3}} \, \tilde\alpha_{\rm max}^{2/3} \,
{\rm e}^{i {\pi \over 3}}
\end{equation}
and
\begin{equation}
\label{G1}
\Gamma_{1,n}  = 3 i k'(\theta') \left[ n + 1/2 \right], \,\,\,\,\,
n=0,1,2,...,
\end{equation}
where $\theta'$ is the matching point which is the point of maximum
of function $\tilde\alpha(\theta)$.  The leading mode corresponds to
the value $n=0$.  Function $\sigma(\theta)$ in Eq.(\ref{n0}) has the
following explicit form:
\begin{equation}
\label{sigma}
\sigma_n(\theta) = {\rm exp} \left\{
{\displaystyle
{\int} \frac{\Gamma_{1,n} - i k' \left( 1 +
\frac{\displaystyle 2 k^2}{\displaystyle \Gamma_0 +k^2} \right)}
{2 i k + \frac{\displaystyle \hat\alpha}
{\displaystyle 2 (\Gamma_0+k^2)}} {d \theta}
}
\right\} .
\end{equation}
Equations (\ref{as}---\ref{sigma}) determine the asymptotic solution
completely. We use these explicit formulae throughout the paper.

\section{Equatorial region}

The crucial assumption of the WKB approximation is that the amplitude
of the eigenvector varies in space much slower than the exponential
factor [see Eq.(\ref{as})]. This implies:
$$
{d \vec n_0 \over d \theta} \ll {1 \over \varepsilon} \vec n_0.
$$
One can easily check that in the vicinity of the equator $\theta
\ll 1$, the following estimate holds:
$$
{d \vec n_0 \over d\theta} \sim {1 \over \theta} \vec n_0.
$$
Thus, in domain $\theta < \varepsilon$, the applicability of the WKB
approximation becomes questionable.

Note, that at $\theta=0$, two different branches $k(\theta)$ merge.  In
the WKB theory, such a point is called ``turning
point''. It is more rule than exception that a WKB solution diverges
at a turning point. Using explicit expressions
(\ref{as}---\ref{sigma}) we, indeed, observe that out solution diverges
at $\theta = 0$. It is possible to show that
$$
\sigma(\theta) \approx \theta^{-1/4},\,\,\,\,\, \mbox{for}\,\,\,
\theta\to 0.
$$
Moreover, one can see that $k'(\theta) \sim \theta^{-1/2}$ diverges
as well.

Note, that these infinities have no physical meaning and appear due to
unjustified approximations. The true solution is indeed a smooth
function everywhere including the equator. This can be easily seen by
expanding equations (\ref{eqA}, \ref{eqB}) in the vicinity of $\theta =
0$:
\begin{equation}
\gamma A(\theta) = \alpha'(0) \theta B(\theta) + A''(\theta),
\label{eqA1}
\end{equation}
\begin{equation}
\gamma B(\theta) =- D A'(\theta) + B''(\theta).
\label{eqB1}
\end{equation}
In these equations $\gamma$ plays the role of an independent external
parameter. In the region $\theta < \varepsilon$, the solution can be
written as a superposition of two waves
\begin{equation}
\label{sol0}
{A(\theta) \choose B(\theta)} =
{A_1 + \delta A_1(\theta) \choose B_1 + \delta B_1(\theta)} {\rm e}^{-\sqrt{\gamma} \theta } +
 {A_2 + \delta A_2(\theta) \choose B_2 +\delta B_2(\theta)} {\rm e}^{\sqrt{\gamma} \theta },
 \end{equation}
 where $A_{1,2}$ and $B_{1,2}$ are some constants to be determined by
 matching of (\ref{sol0}) with the solution in the main domain.  The
 $\theta$-dependent corrections to the amplitudes in Eq.(\ref{sol0})
 can be easily found explicitly.  It is possible to show, that for
 $\theta \,{}_{\displaystyle \sim}^{\displaystyle <}\, \varepsilon$ these
 corrections are small and can be safely neglected in the leading
 approximation.  Solution (\ref{sol0}) is a perfectly smooth
 function.

We denote the solution in the northern hemisphere as $\vec n(\theta)$
and in the southern hemisphere as $\vec s(\theta)$.
Let us note, that the first term in Eq.(\ref{sol0})
 describes a plane wave propagating towards the equator, while the
 second one describes the wave propagating polewards and decaying
 exponentially.  The first term can be matched with the WKB solution
 in the northern hemisphere $\vec n(\theta)$.
The only way to match
 the second one is to suppose that the wave generated in the southern
 hemisphere $\vec s(\theta)$ penetrates to the northern hemisphere.
 In the vicinity of the equator the exponential factors in
 Eq.(\ref{sol0}) and in the WKB solution coincide. Equating the WKB
 eigenvector $\vec n_0$ with the first exponential term coefficient in
 (\ref{sol0}) and $\vec s_0$ with the second one, we can express the
 coefficients $A_{1,2}$ and $B_{1,2}$ through the WKB parameters.

Note that there are four different branches of roots $k(\theta)$ of
 Eq.(\ref{HJ}). Two of them describe the dynamo wave in the main
 domain. As it was shown by Galitski and Sokoloff (1999), the third
 branch describes the wave reflected from the pole. One can check that
 the fourth branch left is the only possible candidate to describe the
 wave propagating over the equator. Only this branch decays all the
 way down to $\theta=-\pi/2$. 

Using the asymptotic expansions in the
 WKB solution for the case $\theta \ll 1$, we obtain:
\begin{equation}
\label{nas}
\vec n =
{ a\,
{\rm e}^{(i \pi/12)}\, \theta_1^{(1/4)}
\choose  \varepsilon^{-2} b\, {\rm e}^{(i \pi /6)}\, \theta_1^{(-1/4)}}\,
\exp\left( -\sqrt{\gamma} \theta \right)
\end{equation}
and the adjoint solution in the southern hemisphere:
\begin{equation}
\label{sas}
\vec s^a(\theta) =
{ \varepsilon^{-2} b\, {\rm e}^{(i \pi /12)}\, \theta_1^{(-1/4)} \choose
a\, {\rm e}^{(-i \pi/3)}\, \theta_1^{(1/4)}}
\exp\left( -\sqrt{\gamma^*} \theta \right),
\end{equation}
where $\theta_1\sim \varepsilon$ is the matching point and the
following real constants have been introduced for brevity:
\begin{equation}
\label{a}
a = 2^{4/3} 3^{1/4} \sqrt{\tilde\alpha'(0)} \tilde\alpha^{2/3}_{\rm
  max}
\end{equation}
and
\begin{equation}
\label{b}
b = 3^{1/2} 2^{(-4/3)}\, \tilde\alpha^{1/3}_{\rm max}
\end{equation}

Now, we can estimate the decrement of spatial decay of the dynamo wave
straying into the other hemisphere.
Indeed, the characteristic latitude of its propagation ahead
is $1/\sqrt{{\rm Re}\gamma}$. For the case $D=-10^3$,
or $\varepsilon=0.1$, we can estimate its leading order approximation
using formulae (\ref{gamma})
and (\ref{G0}) as 0.28, or $16^\circ$.
Notice that for $D=-10^4$ this value is of order $7^\circ$.
These numbers look quite reasonable in view of the 
observational data obtained by Harvey (1992).
She found the reversed polarity active regions
that violate Hale's polarity law. Such regions appear in the final phase
of the solar cycle mainly at low latitudes, near the solar equator.
These results give signatures of the straying dynamo wave
propagating from the counterpart hemisphere.

\section{Symmetry properties}

To formalize the subsequent calculations, let us introduce the
following symmetry operator $\hat P$ which we define as follows:
$$
\hat P f(\theta) = f(-\theta).
$$
This operator has the following obvious eigenvalues (parity):
$$
p= \pm 1.
$$
Acting by this operator on the both sides of Eq.({\ref{MFE}) and
  taking into account that $\hat P \alpha({\bf r}) = - \alpha({\bf
    r})$ and $\hat P {\bf v}({\bf r}) = {\bf v} ({\bf r})$,
 we see
  that the operator on the right-hand side commutes with $\hat P$.
  Thus, its eigenfunctions, {\em i.e.} magnetic field ${\bf B}$, can
  be classified by the parity index $p$.  Value $p=+1$ corresponds to
  the quadrupole solution, while value $p = -1$ corresponds to the
  dipole one.  Eigenvalues $\gamma$ corresponding to different
  parities do not coincide.

  Since, Eqs.(\ref{eqA}.\ref{eqB}) [or equivalently Eq.(\ref{spform})]
  follow from Eq.(\ref{MFE}), they inherent the symmetry properties of
  the initial equations. Let us note that these reduced
  equations involve a toroidal component of the physical magnetic
  field and a component of the gauge field.  These two fields have
opposite parities.  Thus,
  when dealing with composite objects like (\ref{vech}) which contain
  the both fields, it is necessary to take into account this
  difference. Let us introduce the following unitary matrix:
\begin{equation}
\label{U}
\hat U = \hat U^\dagger =
\left(
\begin{array}{cc}
+1 &
0 \\
0 & -1
\end{array}
\right).
\end{equation}
The dipole solution satisfies the following condition
\begin{equation}
\label{d}
\hat P \hat U \vec d = \vec d
\end{equation}
and consequently the quadrupole one is defined by
\begin{equation}
\label{q}
\hat P \hat U \vec q = -\vec q.
\end{equation}
The corresponding equations are written as
\begin{equation}
\label{eqd}
\hat {\cal H}  \vec d = \gamma_d\, \vec d\,\,\,\,\, \mbox{and} \,\,\,\,\, \hat H \vec q = \gamma_q\, \vec q.
\end{equation}
We also define the eigenvectors for the adjoint operator (\ref{Hd}) as
\begin{equation}
\label{eqdx}
\hat {\cal H}^{\dagger}  \vec d^a = \gamma_d^*\, \vec d^a\,\,\,\,\, \mbox{and} \,\,\,\,\,
\hat {\cal H}^{\dagger}  \vec q^a = \gamma_q^* \vec q^a.
\end{equation}
Since the operator is nonhermitian, the eigenvectors for the mutually
adjoint operators do not coincide.
Let us note here that the eigenvalues may
exist only in the form of complex conjugated pairs. Formally,
this follows from the fact that operator $\hat{\cal H}$ as well as the
magnetic field are real.
Below we are mainly interested in the structure of eigenfunctions.
However, the splitting of the eigenvalues is of some interest as well.

To find the splitting of the eigenvalues, we make use of a method used
to solve a well-known quantum-mechanical problem. Namely, we observe
that the problem at hand is very similar to the double-well problem in
quantum mechanics.  In the latter, a particle in a symmetric
one-dimensional potential is studied.  The potential consists of two
quantum wells separated by a high barrier. If one neglects the
possibility of the penetration through the barrier the eigenvalues are
degenerated and they can be calculated in a well (say in the right
one) with the help of the WKB approximation. The eigenfunctions (wave
functions) decay exponentially far from the well.  If one takes into
account a finite probability of the barrier penetration, the
degeneracy is lifted and the energy levels split into pairs
corresponding to the symmetric and antisymmetric solutions. The
quantity of interest is the energy difference between the two lowest
eigenstates, which corresponds to the tunneling rate through the
double well barrier. The quantity is often small and difficult to
calculate numerically, especially when the potential barrier between
the two wells is large.  However, using the WKB eigenfunctions
obtained for each quantum wells, it is
easy to construct approximate symmetric and antisymmetric solutions
explicitly. The subsequent calculation of the tunneling rate is
straightforward and simple (see {\em e.g.} Landau and Lifshitz, 1968).

In our problem we are dealing with the southern and northern domains
of generation separated by the equatorial region in which magnetic
field generation is weak. This equatorial region corresponds to the
barrier.  Neglecting the interaction of the dynamo waves generated in
different hemispheres, one can obtain the WKB eigenfunctions and
eigenvalues explicitly [see Eqs.(\ref{as}---\ref{sigma})].  Using the
solution in the northern hemisphere $\vec n(\theta)$ and in the
southern one $\vec s(\theta)$, we follow the classical Lifshitz
solution of the quantum problem and construct the dipole and
quadrupole eigenfunctions as follows
\begin{equation}
\label{dd}
\vec d(\theta) =  {1 \over \sqrt{2}}\, \left[ \vec n(\theta) + \vec s(\theta) \right]
\end{equation}
and
\begin{equation}
\label{qq}
\vec q(\theta) =  {1 \over \sqrt{2}}\, \left[ \vec n(\theta) - \vec s(\theta) \right],
\end{equation}
where factor $1/\sqrt{2}$ is introduced in order to preserve the norms of the
eigenvectors.

Let us note, that the solution in the southern hemisphere can be
obtained easily by writing
$$
\vec s(\theta) = \hat P \hat U \vec n(\theta).
$$

To proceed further, let us introduce the following ``inner product''
of two vector functions $\vec f$ and $\vec g$:
\begin{equation}
\label{scal}
\left( \vec f, \vec g \right) = \int\limits_{0}^{\pi/2}\,
\left[ f_1(\theta) g_1^*(\theta) + f_2(\theta) g_2^*(\theta) \right]
\cos{\theta} d\theta,
\end{equation}
where
$$
\vec f = {f_1 \choose f_2}\,\,\,\,\, \mbox{and} \,\,\,\,\,\vec g =
{g_1 \choose g_2}.
$$
By multiplying equation $\hat {\cal H} \vec n = \gamma_0 \vec n$ on
$\vec d^a$ and Eq.(\ref{eqdx}) on $\vec n$, we obtain
\begin{equation}
\label{1}
\gamma^*_d - \gamma_0 =
\frac
{\left(\left(\hat {\cal H}^{\dagger} \vec d^a\right)^*\, ,\, \vec n \right) -
\left( \vec d^{a*}\, ,\, \hat {\cal H} \vec n \right)}
{\left( \vec d^{a*}\, ,\, \vec n \right)}
\end{equation}
and
\begin{equation}
\label{2}
\gamma^*_q - \gamma_0 =
\frac
{\left(\left( \hat {\cal H}^{\dagger} \vec q^a\right)^*\, ,\, \vec n \right) -
\left( \vec q^{a*}\, ,\, \hat {\cal H} \vec n \right)}
{\left( \vec q^{a*}\, ,\, \vec n \right)}.
\end{equation}
Let us note here that the dynamo-wave generated in the southern hemisphere
$\vec s(\theta)$, if exists, is exponentially small in the northern
one. Thus, we conclude:
$$
\left| \left( \vec s^{a*}\, ,\, \vec n \right) \right| \ll \left|
  \left( \vec n^{a*}\, ,\, \vec n \right) \right|.
$$
Hence
$$
\left( \vec d^{a*}\, ,\, \vec n \right) \sim \left( \vec q^{a*}\, ,\,
  \vec n \right) \sim \left( \vec n^{a*}\, ,\, \vec n \right)
$$
With a good accuracy, we can neglect the corresponding difference
in the denominators of expressions (\ref{1}) and (\ref{2}) and obtain
the following important identity:
\begin{equation}
\label{dgamma}
\gamma^*_d - \gamma_q^* = 2
\frac
{\left(\left(\hat {\cal H}^{\dagger} \vec s^a\right)^*\, ,\, \vec n \right) -
\left( \vec s^a\, ,\, \hat {\cal H} \vec n \right)}
{\left( \vec n^{a*}\, ,\, \vec n \right)}.
\end{equation}
This equation brings the dynamo problem into direct correspondence
with the quantum one.

Using the explicit expressions for the operators $\hat {\cal H}$ and
$\hat {\cal H}^\dagger$, we obtain the following relation:
\begin{eqnarray}
\label{int}
{1 \over 2} {\left( \vec n^{a*}\, ,\, \vec n \right)}\, \Delta \gamma^*  =
\int \limits_0^{\pi /2}
d\theta \cos{\theta} \Bigl[ \Delta s_1^a  n_1 + D {d \over d\theta}
\left( \cos{\theta}\, s_2^a \right) n_1 + \Delta s_2^a  n_2 \nonumber \\
-
 s_1^a  \left( \Delta n_1 \right)
+ D  s_2^a  {d \over d\theta}
\left( \cos{\theta}\, n_1 \right) -
 s_1^a  \left( \Delta n_2 \right) \Bigr],
\end{eqnarray}
where indexes ``1'' and ``2'' correspond to the upper and lower
components of a vector and $s^a$ means the solution of the adjoint
equation in the southern hemisphere. Integral Eq.(\ref{int}) can be
easily evaluated by parts and the splitting is expressed as boundary
terms.
\begin{equation}
\label{boun}
{1 \over 2} {\left( \vec n^{a*}\, ,\, \vec n \right)}\, \Delta \gamma^*  =
n_1 {d s_1^a \over d \theta}
+ n_2  {d s_2^a \over d \theta}
-  s_1^a  {d n_1 \over d \theta}
-  s_2^a  {d n_2 \over d \theta}
+ D n_1 s_2^a   \Biggr|_{\theta=0}
\end{equation}

\section{Eigenvalue splitting}

In this section we calculate the eigenvalue splitting explicitly using
the WKB solution obtained previously (see Sec. 2). First of all, we
find the eigensolution of the adjoint operator (\ref{Hd}).

Let vector  $\vec n(\theta)$, with the upper component $n_1(\theta)$
and lower $n_2(\theta)$, be the WKB solution of Eq.(\ref{spform})}
[see Eqs. (\ref{as}---\ref{sigma})].
Then, the corresponding solution of the adjoint equation can be
obtained by the following replacements:
$$
n_{1,2} \to n_{2,1},
$$
$$
\Gamma_0 \to \Gamma_0^*\,\,\,\,\, \mbox{and} \,\,\,\,\, \Gamma_1
\to \Gamma_1^*.
$$
Function $k(\theta)$ satisfying the Hamilton-Jakobi equation
(\ref{HJ}) should be replaced by the following value:
$$
k(\theta) \to -k^*(\theta).
$$
After the appropriate replacements are made, the adjoint solution
takes the form:
\begin{equation}
\label{nna}
\vec n^a  (\theta) = {-i \varepsilon^{-2} k^*(\theta) \cos{\theta} \choose
\Gamma_0^* + k^*(\theta)^2} {\sigma^*(\theta) \over \cos{\theta}}
{\rm exp} \left[ - {i S^*(\theta) \over \varepsilon}\right],
\end{equation}

At this point, we can calculate the eigenvalue splitting using the
explicit expressions obtained above. We start with the evaluation of
the inner product $\left( \vec n^a\, ,\, \vec n \right)$. Let us note,
that in the double-well problem the corresponding product is nothing
but the norm of the wave-function and, thus, is equal to unity.
Using Eqs.(\ref{n0}) and (\ref{nna}) we get:
\begin{eqnarray}
\left( \vec n^a\, ,\, \vec n \right) =
{1 \over \varepsilon^2} \int \limits_0^{\pi / 2} &d \theta&
\left[ - \left( \Gamma_0 + k^2(\theta) \right) i k^*(\theta) +
i k(\theta) \left( \Gamma_0 + k^2(\theta) \right)^*  \right] \nonumber \\
&\times& \left| \sigma(\theta) \right|^2\,
 \exp\left\{ - {2 \over \varepsilon} \int \limits_0^{\theta}
 {\rm Im\,} k(\theta') d \theta' \right\}.
\end{eqnarray}
Since the integrand in the above expression contains a Gaussian
exponent with a sharp maximum, it is possible to evaluate the integral
by the Laplace method. In the saddle point approximation, we obtain
\begin{equation}
\label{norma}
\left( \vec n^{a*}\, ,\, \vec n \right) =
{4 {\rm Im\,} \Gamma_0 \over \varepsilon^2}
\sqrt{ \pi \varepsilon \over {\rm Im\,} k'(\theta_*) }
\left| \sigma(\theta_*) \right|^2 k(\theta_*)
\exp\left\{ - {2 \over \varepsilon} \int \limits_0^{\theta_*}
 {\rm Im\,} k(\theta) d \theta \right\},
\end{equation}
where $\theta_*$ is the point where action $S(\theta)$ reaches its
maximum, {\em i.e.} the point at which ${\rm Im\,}k(\theta_*) = 0$.
This point has already been found explicitly by Kuzanyan and Sokoloff
(1995) as follows
\begin{equation}
\label{theta1}
{\tilde\alpha(\theta_*) \over \tilde\alpha_{\rm max}} =
{ 9 \sqrt{3} \over 16 \sqrt{2} \sqrt{\sqrt{3} - 1}} \approx 0.8052.
\end{equation}
Let us note that quantity (\ref{norma}) is a real, positive and exponentially
large number.  Thus, the eigenvalue splitting is exponentially small which is
in accord with the familiar result for the double well problem.

%
%
%


Using expressions (\ref{nas})--(\ref{b}),
we obtain the following results for
bo\-un\-dary terms (\ref{boun}):
\begin{equation}
\label{b1}
\left[ n_1 {d s_1^a \over d \theta} - s_2^a {d n_2 \over d
  \theta}\right] \, \Biggl|_{\theta=0}=
{2 a b \over \varepsilon^3} {\rm Re\,}\left[ {\rm e}^{(-i \pi /6)}
  \sqrt{\Gamma_0} \right],
\end{equation}
\begin{equation}
\label{b2}
\left[ n_2 {d s_2^a \over d \theta} - s_1^a {d n_1 \over d
  \theta} \right] \, \Biggl|_{\theta=0}=
{2 a b \over \varepsilon^3} {\rm Re\,}\left[ {\rm e}^{(-i \pi /6)}
  \sqrt{\Gamma_0} \right],
\end{equation}
\begin{equation}
\label{b3}
  D n_1 s_2^a\, \Bigr|_{\theta=0}=
-{a^2 \sqrt{\theta_1} \over \varepsilon^3}
{\rm e}^{(-i\pi/3)}.
\end{equation}
The two first boundary terms give real positive contributions to the
splitting. This is not very surprising, since the corresponding terms
come from the hermitian part of operator $\hat {\cal H}$.  What is more
remarkable is that the matching point $\theta_1$ drops out of the
final result for these terms.  The third ``nonhermitian'' term gives a
nonvanishing contribution to the imaginary part of the eigenvalue
difference and explicitly depends upon the matching point.
Since $\theta_1 \sim
\varepsilon$, we see that this contribution is parametrically small
compared to the first two ones.

Using Eqs.(\ref{b1}, \ref{b2}, \ref{b3}) we obtain the following result:
\begin{equation}
\label{finRe}
\left(\vec n^a\,,\,\vec n \right) {\rm Re\,}\Delta\gamma =
{1 \over \varepsilon^3}\,
{ 3^{9/4} \over 2^{1/3} }\, \tilde\alpha_{\rm max}^{4/3}\,
\sqrt{\tilde\alpha'(0)}
\end{equation}
and
\begin{equation}
\label{finIm}
\left(\vec n^a\,,\,\vec n \right) {\rm Im\,}\Delta\gamma =
- {\sqrt{\varepsilon}\over {\varepsilon^3}}
\, 2^{19/6} 3^{1/2} \sqrt{\theta_1 \over \varepsilon} \,
\tilde\alpha_{\rm max}^{4/3}\, \tilde\alpha'(0),
\end{equation}
where norm $\left(\vec n^{a*}\,,\,\vec n \right)$ was calculated in
Sec.~4 [see Eq.~(\ref{norma})]. The matching point $\theta_1$
can be calculated
self-consistently as a point at which the phase-shifts for different
asymptotics coincide [{\em cf.} Galitski and Sokoloff (1999)].

As we have seen above, the WKB approximation
breaks down near the equator. This happens because $\theta=0$ is the
turning point for our solution. Thus, straightforward evaluation of
(\ref{boun}) using the WKB formulae is not possible.  Note that in
the usual quantum problem such a difficulty does not arise since the
boundary is located far from the turning point. The WKB solution can
be applied directly and the exponential term coefficient can be found
easily.
However, it is important to realize that the method itself is limited
by the exponential accuracy. The numerical part of the
exponential term coefficient should not be trusted even if found. There
are other
asymptotical methods allowing the calculation of the tunneling rate,
which give slightly different results. Indeed, the exponential factor
is the same for all these methods. For a detailed discussion of the
asymptotical methods in quantum double-well problems see {\em e.g.}
Cooper {\em et al.} (1995).
%
However, as we have already mentioned, the method used does not
allow us to obtain the correct value of the exponential term
coefficient.
We conclude that the accuracy of Eqs.(\ref{finRe}, \ref{finIm})
already exceeds the accuracy of the method. Hence, the exact value of
$\theta_1 \sim \varepsilon$ is not important.

Bearing these observations in mind, let us write the final result in
the following form:
\begin{equation}
\label{finalRe}
{{\rm Re\,}\Delta\gamma \over \left| \gamma_0 \right|} \approx
\sqrt{\varepsilon}
 \exp\left\{ {2 \over \varepsilon} \int_0^{\theta_*}
\left| {\rm Im\,} k(\theta)\right| d \theta \right\}
\end{equation}
and
\begin{equation}
\label{finalIm}
{{\rm Im\,}\Delta\gamma \over \left| \gamma_0 \right|} \approx
-\varepsilon
\exp\left\{- {2 \over \varepsilon} \int_0^{\theta_*}
\left| {\rm Im\,} k(\theta)\right| d \theta \right\}
\end{equation}

\section{Discussion}

The overall picture of the magnetic field in the Parker model can be
now summarized as follows. A strong dynamo wave appears at middle
latitudes. The point where the generation sources reach maximum is
shifted from the point where the helicity coefficient is maximal. This
happens because the differential rotation and helicity mechanisms are
both important on equal footing.  In the framework of the asymptotic
analysis, the two mechanisms manifest themselves through a universal
function. The asymptotic WKB theory allows to derive explicit
expressions describing the generated dynamo waves. It appears that at
middle latitudes the main wave propagates equatorwards. At very high
latitudes the wave changes its direction of motion and propagates
towards the pole and reflects from it.  In the vicinity of the equator
there is a transition zone $\theta \sim \epsilon$ where the WKB
solution becomes progressively worse and the crossover from the
asymptotic behaviour (\ref{as}) to (\ref{sol0}) occurs.  The wave does
not vanish completely at the equator but rolls over it propagating in
the southern part. Beyond the southern transition zone $\theta \sim -
\varepsilon$, the WKB solution becomes applicable again and it
describes a very weak decaying dynamo wave.

Let us mention here the observational results which refer to the time
of the Maunder minimum (see Sokoloff and Nesme-Ribes, 1994;
Ribes and Nesme-Ribes, 1993). During
the Maunder minimum, the solar magnetic activity was very weak.
At the end of the
minimum, some signs of the activity appeared in the southern
hemisphere. The southern wave of magnetic activity looked normal and
propagated equatorwards. In the northern part there was almost no
activity at all, except an unusual wave which existed only near the
equator and propagated away from it.

Certainly, our simple linear asymptotic theory can not suggest any
explanations for the Maunder minimum itself. However, we think that
the weak equatorial wave detected at the epoch of the  minimum
corresponds to the equatorial wave which appears naturally in the
framework of our asymptotic analysis. On the modern Sun, this
equatorial wave, if exists, is screened by the background of the
main dynamo wave.
There are observational signatures of some magnetic field
tracers such as extreme ultraviolet (EUV) lines to propagate polewards
near the solar equator
({\em e.g.}, Benevolenskaya et al., 2001).
They may indicate the presence of such a straying wave propagating
from the other hemisphere.
Notice, that the analysis of the observational data by Harvey (1992)
revealed that the reversed polarity active regions
that violate the Hale's polarity law are located mainly at low latitudes.
Further observations of the equatorial
region would be highly desirable to clarify this question.

We have shown that a small interaction between the northern and
southern waves yields a splitting of the eigenvalues corresponding to
the dipole and quadrupole configurations.  It appears that the growth
rate corresponding to the dipole configuration exceeds the quadrupole
one.  This fact gives a very tentative indication that the dipole
configuration is to dominate the quadrupole one after a sufficiently long
time. Let us estimate this time using the following trial parameters:
$\alpha(\theta)=\sin{\theta}$ and $D=-10^3$. In this case
$$
{{\rm Re\,}{\Delta \gamma} \over \left| \gamma_0 \right|} \sim 0.03
$$
If we suppose that $\left( 2 \pi /{\rm Im\,}\gamma_0 \right) \approx
22\,\mbox{yr}$, where ${\rm Im\,}\gamma_0 =|\gamma_0|\sqrt{3}/2$.
Then we have the following estimate for the time of dominance of the
dipole mode over the quadrupole one $\tau=1/{\rm Re\,}{\Delta \gamma}
 \sim 100\, \mbox{yr}$.
This is comparable with the time of recovery of the solar magnetic field
generation (Ribes and Nesme-Ribes, 1993) from the Maunder Minumum
when some activity was confined to one southern hemisphere (equatorial
asymmetry).

The difference in the imaginary part of the eigenvalues may give
rise to an additional weak period of oscillations of the magnetic
field in the non-linear regime. The corresponding estimate for the
period of such a global modulation is
$T_2=2 \pi /{\rm Im\,}{\Delta \gamma} \sim 2\cdot 10^3\,\mbox{yr}$.

Let us stress that the applicability of the simple Parkler's
migratory dynamo model is limited.
A two-dimensional and non-linear
generalizations are required to provide a more accurate description.
However, some simple estimates can be done now.
Namely, using
a two-dimensional asymptotic solution, {\em e.g.}, obtained recently by
Belvedere {\em et al.} (2000), we can estimate the overlap of the dynamo
waves generated in different hemispheres.
Because the r.h.s. of equation (\ref{int}) then turns to a 2D
integral, we expect that the corresponding overlap exceeds significantly
that one in the one-dimensional model.
This may result in a larger splitting of the eigenvalues. The
corresponding period of secondary oscillations $T_2$ should increase.
This qualitative result suggests a possible explanation for the
well-known Gleissberg cycle.

%

\noindent
{\it Acknowledgements}

{\footnotesize\noindent
The work of V.G. was supported by NSF Grant DMR-9812340.
K.K. would like to acknowledge support {}from the
RFBR under grants 00-02-17854, and 99-02-18346a,
the Young Researchers' grant of Russian Academy of Sciences,
NATO grant PST.CLG.976557 and INTAS grant 99-00348.
}
$$$$

\noindent
{\it References}

Bassom, A. P., Kuzanyan, K. M., and Soward, A. M., ``A nonlinear
dynamo wave riding on a spatially varying background.''
{\it Proc. R. Soc. Lond.} A, {\bf 455}, p. 1443 (1999).

Belvedere, G., Kuzanyan, K. M., and Sokoloff, D., ``A two-dimensional
asymptotic solution for a dynamo wave in the light of the solar
internal rotation,'' {\it Mont. Not. Royal Astron.  Society}, {\bf
  315}, 778-790 (2000).

Benevolenskaya, E.E., Kosovichev, A.A., Schereer, P.H.,
``Detection of High-Latitude Waves of Solar Coronal Activity in
Extreme-Ultraviolet Data From the Solar and Heliospheric Observatory
EUV Imaging Telescope''
{\it Astrophys. J. Lett.} {\bf 554}, {L107-L110} (2001)

Cooper, F., Khare, A., and  Sukhatme, U.,
``Supersymmetry and Quantum Mechanics,''
{\it Phys.Rept.} {\bf 251} p. 267 (1995).

Harvey, K.L.
``The Cyclic Behavior of Solar Activity"
in: K.L. Harvey (ed.) {\it The Solar Cycle, ASP Conference Series}
{\bf 27}, {335--367}
(1992)

Galitski, V. M. and Sokoloff, D. D., ``Spectrum of Parker's
Equations,'' {\it Astron. Rep.} {\bf 42}, p. 127 (1998).

Galitski, V. M. and Sokoloff, D. D.,
``Kinematic dynamo wave in the vicinity of the solar poles,''
{\it Geophys. Astrophys. Fluid Dynamics}, {\bf 91}, p. 147 (1999).

Krause, F. and R\"adler, K.-H., {\it Mean-Field Electrodynamics and
Dynamo Theory}, Oxford: Pergamon Press (1980).

Kuzanyan, K. M. and Sokoloff, D. D., ``A dynamo wave in an
inhomogeneous medium,'' {\it Geophys. Astrophys. Fluid Dyn.} {\bf 81},
p. 113 (1995)

Kuzanyan, K. M. and Sokoloff, D. D., ``A
dynamo wave in a thin shell,'' {\it Astron. Rep.} {\bf 40}, p.
425 (1996).

Landau, L. D. and Lifshitz, E. M., {\it Quantum Mecahnics},
Addison-Wesley, Mass. (1958).

Makarov, V. I. and Sivaraman, K. R. ``Poleward migration of the
magnetic neutral line and reversals of the polar fields on the
Sun,'' {\it Sol. Phys.} {\bf 85}, p. 215 (1983).

Maslov, V. P. and Fedorjuk, M. V. {\it Semi-Classical Approximation
in Quantum Mechanics}, D. Reidel: Dordrecht (1981).

Meunier, N., Proctor, M. R. E., Sokoloff, D. D., Soward, A. M. and
Tobias, S. M., ``Asymptotic properties of a nonlinear
$\alpha\omega$-dynamo wave: period, amplitude and latitude
dependence.'' {\it Geophys. Astrophys. Fluid Dynam.} {\bf 86}, p. 249
(1997).

Parker, E. N., ``Hydromagnetic dynamo models,'' {\it Astrophys. J.}
{\bf 122}, p. 293, (1955).

Ribes J.C., and Nesme-Ribes, E., ``The Solar Sunspot Cycle in the Maunder
Minumum AD 1645 to Ad 1715,
{\it Astron. Astrophys.} {\bf 276}, {549--563}
(1993)

Ruzmaikin, A., Shukurov, A., Sokoloff, D., Starchenko, S.,
\lq \lq Maximally-Efficient-Generation Approach in the Dynamo
Theory", {\it Geophys. Astrophys. Fluid Dyn.} {\bf 52}, 125 - 139 (1990)

Schou J.%
, Antia H.M., Basu S., Bogart R.S., Bush R.I., Chitre S.M.,
Christensen-Dalsgaard J., Di Mauro M.P., Dziembowski W.A.,
Eff-Darwich A., Gough D.O. , Haber D.A., Hoeksema J.T., Howe R.,
Korzennik S.G., Kosovichev A.G., Larsen R.M., Pijpers F.P.,
Scherrer P.H., Sekii T.,  Tarbell T.D., Title A.M., Thompson
M.J., Toomre J.,
1998,
Helioseismic Studies of Differential Rotation in the Solar
 Envelope by the solar Oscillations Investigation using the
Michelson Doppler Imager.
{\it Astrophys. J.}
{
\bf
505}, {390%
--417
}

Steenbeck, M., Krause, F., and R\"adler, K.-H., ``Berechnung der
mittlere Lorentz-FeldSt\"arke f\"ur ein elektrisch leitendes Medium in
turbulenter, durch Coriolis-Kr\"afte beeinflusster Bewegung,'' {\it Z.
  Naturforsch}, {\bf 21}, p. 369 (1966).

Sokoloff D., Shukurov A., and Ruzmaikin A., \lq \lq Asymptotic Solution of
the $\alpha^2$-Dynamo Problem", {\it Geophys. Astrophys. Fluid Dynam.}
{\bf 25}, 293 -- 307 (1983)

Sokoloff D.D., Nesme-Ribes, E. \lq \lq The Maunder Minimum, a
Mixed-Parity Dynamo Mode?"
{\it Astron. Astrophys.} {\bf 288}, 293 -- 298 (1994)

\end{document}